\documentclass[doublecol]{epl2}

\def\cm{cm$^{-1}$}
\def\TCO{$T_{\rm CO}$}
\def\LCMO{La$_{0.25}$Ca$_{0.75}$MnO$_3$}

\title{Origin of Low-Energy Excitations in Charge-Ordered Manganites 
}
\author{E. Zhukova\inst{1,2} \and B. Gorshunov\inst{1,2} \and T. Zhang\inst{1}
\thanks{E-mail: \email{zhangtao@issp.ac.cn}}
\and Dan Wu\inst{1} \and A.S. Prokhorov\inst{2} \and \\ V.I.
Torgashev\inst{3} \and E.G. Maksimov\inst{4} \and M.
Dressel\inst{1}
\thanks{E-mail: \email{dressel@pi1.physik.uni-stuttgart.de}}
}
\institute{
  \inst{1} 1.~Physikalisches Institut, Universit{\"a}t
Stuttgart, Pfaffenwaldring 57, 70550 Stuttgart, Germany\\
  \inst{2} Prokhorov Institute of General Physics, Russian Academy of Sciences, Vavilova Str. 38, 119991 Moscow, Russia\\
  \inst{3} Faculty of Physics, Southern Federal University, Zorge Str. 5, 344090 Rostov-on-Don, Russia\\
  \inst{4} Lebedev Physics Institute, Russian Academy of Sciences, Leninsky prosp., 53, 119991, Moscow, Russia
 }
\shortauthor{E. Zhukova, B. Gorshunov, T. Zhang {\it et al.}}

\abstract{The low-energy  excitations in the charge-ordered phase
of polycrystalline La$_{0.25}$Ca$_{0.75}$MnO$_3$ are explored by
frequency-domain terahertz spectroscopy. In the frequency range
from 4~\cm\ to  700~\cm\ (energies 0.4~meV to 90~meV) and at
temperatures down to 5~K, we do not detect any feature that can be
associated with the collective response of the spatially modulated
charge continuum. In the antiferromagnetically ordered phase,
broad absorption bands appear in the conductivity and permittivity
spectra around 30~\cm\ and 100~\cm\ which are assigned to former
acoustic phonons optically activated  due to a fourfold
superstructure in the crystal lattice. Our results indicate that
characteristic energies of collective excitations of the
charge-ordered phase in La$_{0.25}$Ca$_{0.75}$MnO$_3$, if any, lie
below 1~meV. At our lowest frequencies of only few wavenumbers a
strong relaxation is observed above 100~K connected to the
formation of the charge-ordered state. }
\begin{document}

\maketitle

\section{Introduction}
Manganese oxides $R_{1-x}A_x$MnO$_3$ (with $R$ being rare earth
and $A$  alkaline elements) represent a unique playground for
investigating electronic correlation effects in solids because the
competing interactions related to charge, spin, orbital and
lattice degrees of freedom are of comparable strength, resulting
in a rich phase diagram \cite{Rao98,Tokura99,Dagotto01}. In recent
years much effort has been devoted to the intriguing phenomenon of
charge ordering (CO), which for La$_{1-x}$Ca$_x$MnO$_3$ in the
concentrations range $0.5 \leq x \leq 0.85$ takes place at rather
high temperatures. Below $T_{\rm CO}$ the system changes also its
magnetic properties from a paramagnetic to an antiferromagnetic
(AFM) state. It was suggested that the  mixed-valence manganese
ions dissociate into two subsystems with integer valences,
Mn$^{3+}$ and Mn$^{4+}$ which occupy different atomic sites of the
lattice, causing spatial stripes of commensurate doping with a
period being a multiple of the lattice  constant in the $a$
direction \cite{Goodenough55, Chen97,Wang00}. Later it was found,
however, that the charge modulation does not have full amplitude
and that the wavevector ${\bf q}={\bf a}^{*}(1-x)$ is
concentration dependent ({${\bf a}^{*}$ is the reciprocal lattice
vector) and not necessarily tied only to atomic sites
\cite{Coey04,Brey04,Loudon05,Cox06}.

Recently,  it was suggested \cite{Milward05} that the ordering in
manganites can be of Fr\"ohlich-Peierls type  leading to a charge
density wave (CDW) \cite{Gruner88}. Several observations seem to
support this idea. Fermi-surface nesting \cite{Chuang01} was
observed,  together with transport signatures of a CDW, such as
nonlinear dc resistivity and broad-band noise \cite{Wahl03,Cox08}.
Terahertz and infrared (IR) spectroscopic measurements on
Nd$_{1-x}$Sr$_x$MnO$_3$, La$_{1-x}$Ca$_x$MnO$_3$ and
Pr$_{1-x}$Ca$_x$MnO$_3$ have revealed resonances in the
far-infrared range which are interpreted by the authors as
collective modes arising from the CDW condensate
\cite{Kida02,Nucara08}. Although the idea seems intriguing
\cite{Dressel09} and applicable to other charge-ordered systems
such as cuprates \cite{Dumm02,Lucarelli03,Lupi09,Vuletic06} or
organics \cite{Dressel94}, more experiments are needed to verify a
CDW state in charge-ordered manganites \cite{Schmidt08,Fisher09}.
Here we present investigations of the low-energy electrodynamics
of La$_{0.25}$Ca$_{0.75}$MnO$_3$ which clearly identify the
low-frequency excitations simply as former acoustic phonons
optically activated by folding of the Brillouin zone caused by a
superstructure in the crystal lattice.

\section{Experimental}
Polycrystalline samples of \LCMO\ were prepared by sol-gel method
as described in Ref.~\cite{Zhang07} where the grain size can be
selected by annealing at different temperatures. The average
particle diameter was determined by field-emission scanning
microscopy and X-ray using Scherrer formula; the structure of the
samples was characterized by X-ray diffraction. For the optical
measurements, we prepared pellets of 10~mm diameter and
thicknesses varying from approximately 0.1 to 0.6~mm. Optical
studies were performed in the THz range using a coherent-source
spectrometer in a Mach-Zehnder arrangement \cite{Gorshunov05} that
allows us to directly measure the complex conductivity
$\hat{\sigma}(\nu)$ or complex permittivity $\hat{\epsilon}(\nu)$
at frequencies from $\nu=1$ up to 50~\cm, in the temperature
interval from 2 to 300 K and in magnetic field up to 8 T. On the
same pellets of \LCMO\ also the far-infrared reflectivity was
measured up to 700~\cm\ using a Fourier transform infrared
spectrometer in order to obtain  the real parts of the
conductivity $\sigma(\nu)$ and permittivity
$\epsilon^{\prime}(\nu)$ in a wide frequency interval from 4 up to
700~\cm.

\section{Results}
\begin{figure}
    \centering
    \includegraphics[width=\columnwidth]{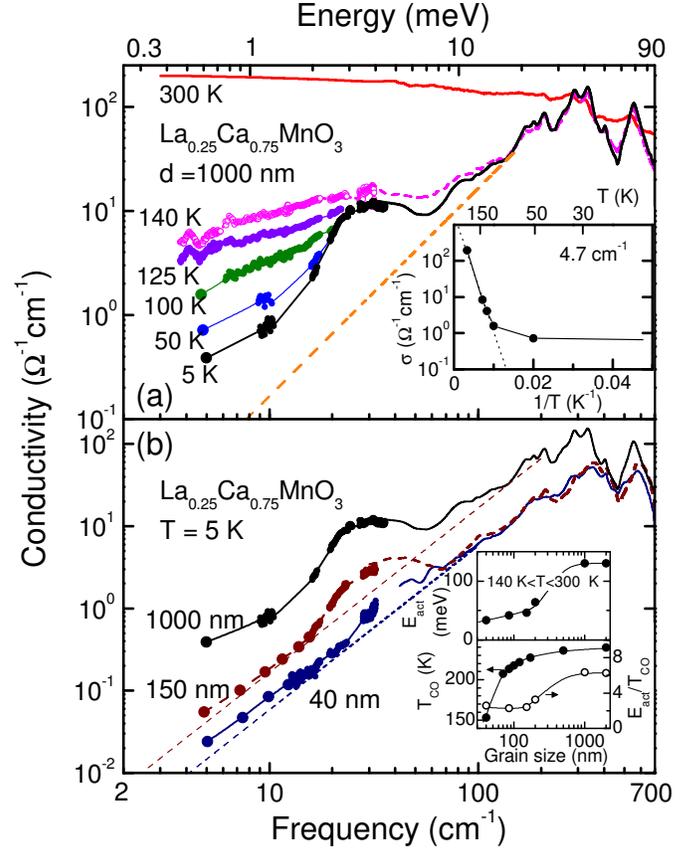}
    \caption{(Color online) (a) Optical conductivity of \LCMO\ plotted in a double-logarithmic
    fashion for various temperatures as indicated.
    The symbols correspond to data directly obtained by a Mach-Zehnder interferometer, solid and dashed lines are from reflectivity measurements; the lines below 40~\cm\ guide the eye. The thin dotted line represents the $\nu^2$ behavior of the conductivity indicating the low-frequency tail of the lowest infrared phonon.
    The inset demonstrates in an Arrhenius plot the activated behavior of the low-frequency ($\nu = 4.7$~\cm) conductivity with an activation energy 0.13~eV (dashed line).
(b) Comparison of the low-temperature conductivity $\sigma(\nu $)
of \LCMO\ samples consisting of different grain sizes, $d=1000$,
150, and 40~nm. The dependence of the activation energy $E_{\rm
act}$, the CO transition temperature $T_{\rm CO}$ and correlation
strength $2\Delta/T_{\rm CO}$ on the grain-size is plotted in the
inset. \label{fig:1}}
\end{figure}
In Fig.~\ref{fig:1}a the conductivity spectra of \LCMO\ pellets
with grains of 1 micrometer  size are plotted for different
temperatures. Above the charge-ordering temperature $T_{\rm
CO}\approx 240$~K the conductivity is metal-like, it is almost
frequency independent below 30~\cm, and gradually decreases for
higher frequencies. Beyond 200~\cm\ phonon features become
dominant because incompletely screened by itinerant carriers.

When the sample is cooled below \TCO, the conductivity spectra
change their character qualitatively, from metallic to dielectric.
The values of $\sigma(\nu)$ drop drastically, especially at the
lowest frequencies of few wavenumbers. The permittivity
$\epsilon^{\prime}(\nu)$ increases from negative values (at
$T=300$~K) up to large positive values; the slope ${\rm
d}\epsilon^{\prime}/{\rm d}\nu$ becomes negative (cf.\
Fig.~\ref{fig:2}a). As demonstrated in the inset of
Fig.~\ref{fig:1}a, in the range $T_N \approx 140~{\rm K}< T <
T_{\rm CO}\approx 240$~K the conductivity is thermally activated
$\sigma(T)\propto\exp\{-E_{\rm act}/k_B T\}$ with  $E_{\rm act} =
0.13$~eV, indicating an energy gap $2\Delta=E_{\rm act}$
(pseudogap for $T > T_{\rm CO}$) in the density of states. It is
generally accepted that this gap or pseudogap is caused  by the
long-range charge order below \TCO\ or by correspondent
order-parameter fluctuations above \TCO\ \cite{Kim02}.

In the AFM phase ($T<T_N$), an absorption band appears in the
range 20 to 40~\cm\, getting more pronounced as the temperature is
lowered. A close inspection of this band reveals two resonances
located at frequencies $25\pm 3$ and $38\pm 3$~\cm, which are
basically temperature independent. When we reduce the grain size
from 1000 to 40~nm, the band gradually disappears, as is
demonstrated by Fig.~\ref{fig:1}b. In addition, a less intense and
broader absorption band can be identified in the range $60 -
100$~\cm. It is also more pronounced in the AFM phase, and looses
intensity in samples with smaller grains. At our lowest
frequencies of only few wavenumbers a strong relaxation is
observed above 100~K which vanishes as the temperature decreases
$T\rightarrow 0$.

\section{Discussion}
It is generally accepted that  the simple picture based on
ordering of Mn$^{3+}$-Mn$^{4+}$ ions is too rough to account for
all peculiarities of manganites. To explain the CO state with
charge modulation of non-integer amplitude and (in general)
incommensurate  wavevector ${\bf q}= {\bf a}^*(1-x)$, a CDW ground
state was proposed \cite{Wahl03,Cox08,Schmidt08}. Kida {\it et
al.} \cite{Kida02} assigned a low-frequency ($16 - 24$~\cm)
absorption peak observed in Pr$_{0.7}$Ca$_{0.3}$MnO$_3$ to a
collective excitation (phason) of the CDW state. Along these
lines, Nucara {\it et al.}\cite{Nucara08} attributed the
asymmetric bands, they recently detected between 15 and 60~\cm\ in
a series of manganites including \LCMO, to a combination of CDW
phason and amplitudon responses, and another  peak  below 10~\cm\
to a pinned CDW-phason.

\begin{figure}
    \centering
    \includegraphics[width=0.94\columnwidth]{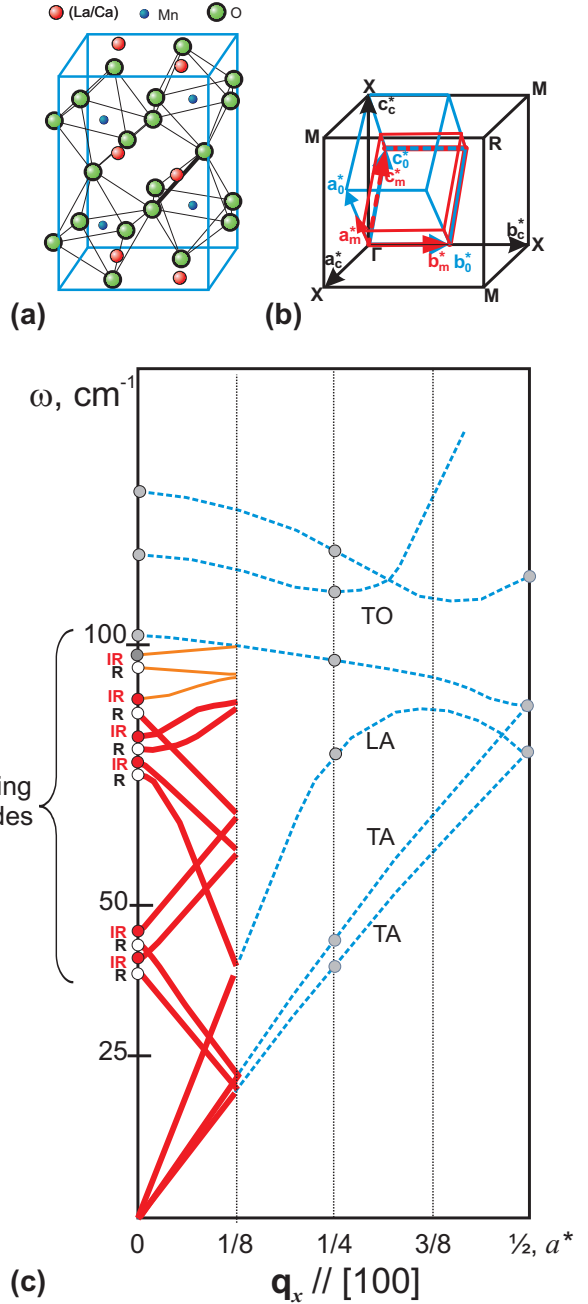}
    \caption{(Color online) (a) Unit cell of (La,Ca)MnO$_3$ in the orthorhombic Pnma phase.
    (b) Transformation of the Brillouin zone of a cubic perovskite-like unit cell at
     the Pm3m$\leftrightarrow$Pnma$\leftrightarrow$P2$_1$/m phase transitions,
      leading to supercell of $4a^*$. (c) Qualitative scheme demonstrating the
       folding of the modes at the Pnma$\leftrightarrow$P2$_1$/m phase transition.
       The acoustic phonons with ${\bf q}_x={\bf a}^*/4$ and ${\bf q}_x={\bf a}^*/2$
       (grey dots) in Pnma structure are folded to the center of the monoclinic zone and become
       Raman- (open dots) and infrared-active (filled dots) optic modes. The band in the
        frequency range around $20 - 40$~\cm\ is formed by modes with wavevector
         ${\bf q}_x= {\bf a}^*/4$, and the band around 100~\cm\ is formed by modes
          with wave vector ${\bf q}_x={\bf a}^*/2$ from the Brillouin zone  boundary.
           Certain contributions to the low-frequency part of the band around 100~\cm\ can
            come from the phonons with the wavevector ${\bf q}_x={\bf a}^*/3$
    (phase with supercell $3a$ [14]).
\label{fig:3}}
\end{figure}

In contrast to these suggestions and based on our comprehensive
spectroscopic investigations of the low-frequency excitations in
\LCMO\, we show that there are no spectroscopic features at
frequencies down to 4~\cm\ which could correspond to a collective
response of the CO-condensate. We assign the resonance absorptions
between 20 and 40~\cm\ and around 100~\cm\ to acoustic phonons
that become optically active after the folding of the Brillouin
zone due to evolution of a superstructure in the crystal lattice
accompanying the charge and antiferromagnetic orderings. Indeed,
X-ray studies of \LCMO\ evidence that the orthorhombic unit cell
quadruples along the $a$ axis when cooled below \TCO\
\cite{Pissas05}, meaning that the high-temperature space group
Pnma is transformed to P2$_1$/m. Consequently, the CO transition
has to be described by an order parameter that is transformed
according to the $\Sigma_1({\bf q}_x={\bf a}^*/4)$ irreducible
representation of the space group Pnma. As a result of the
Pnma~$\leftrightarrow$~P2$_1$/m structural phase transition, the
phonons with the wavevector ${\bf q}_x={\bf a}^*/4$ and ${\bf
q}_x={\bf a}^*/2$ are folded to the Brillouin zone center
($\Gamma$-point) of the monoclinic P2$_1$/m phase. They are split
into two types of phonons -- symmetric (gerade) and anti-symmetric
(ungerade), relative to the center of inversion; in other words,
they are split into the (g,u) pairs. The gerade modes (A$_g$ and
B$_g$), become optically Raman-active phonons, while the ungerade
modes (A$_u$ and B$_u$) form polarization waves and become
IR-active. In the case of \LCMO\ the infrared activity of the
corresponding modes is promoted by a dipole moment produced by
spatial charge disproportionation which has to include all ions of
the new unit cell together with the valence difference on
manganese ions.

The described transformation of phonon branches is qualitatively
depicted in Fig.~\ref{fig:3}, in close resemblance with
calculations of the phonon frequencies of undoped LaMnO$_3$
\cite{Rini07}. It now becomes obvious that the absorption band
observed in \LCMO\ at $20 - 40$~\cm\ corresponds to the modes with
the wave vector  ${\bf q}_x={\bf a}^*/4$, while the broad set of
superimposed bands at frequencies around 100~\cm\ is associated
with the Brillouin-zone-boundary modes of wave vectors ${\bf
q}_x={\bf a}^*/4$. Significant broadening of the phonon modes can
be connected to the polycrystalline nature of the samples with all
three crystallographic directions contributing to the optical
response and with phases of  other symmetries present below $T_N$,
like a phase with the $3a$ superlattice \cite{Pissas05}. The
appearance of the broad band at $20 - 40$~\cm\ only in the AFM
phase and not right below \TCO\ agrees well with the temperature
evolution of the weight fraction of the new $4a$ phase which
starts to increase below \TCO\ but reaches its full strength only
below $T_N$. Note that the magnetic origin of the band can
definitely be ruled out because the AFM resonance modes are weaker
by more than a factor of  100 and their resonance frequencies are
strongly temperature dependent \cite{Ivannikov02}.

As seen from Fig.~\ref{fig:1}b, the phonon bands weaken gradually
in samples with smaller grain size, indicating a suppression of
the $4a$-superlattice in small nano-crystallites. Reduction of
grain sizes causes an intensity decrease, but not a full
suppression of the CO gap $2\Delta$ and correlation strength
$2\Delta/T_{\rm CO}$ (inset in Fig.~\ref{fig:1}b), in accordance
with magnetic \cite{Zhang09} and optical  measurements
\cite{Kim02}, indicating weak coupling of charge and magnetic
order parameters to the lattice. It would be of interest to
investigate the low-frequency response and nano-size effects in
manganites with other alkali-metal concentrations since $x$
determines the stability of and the balance between the different
interactions and also the frequency position of the folded phonon
modes, according to the wave-vector of the lattice modulation
${\bf q}={\bf a}^{*}(1-x)$ \cite{Brey04,Loudon05,Cox08b}.

\begin{figure}
    \centering
    \includegraphics[width=\columnwidth]{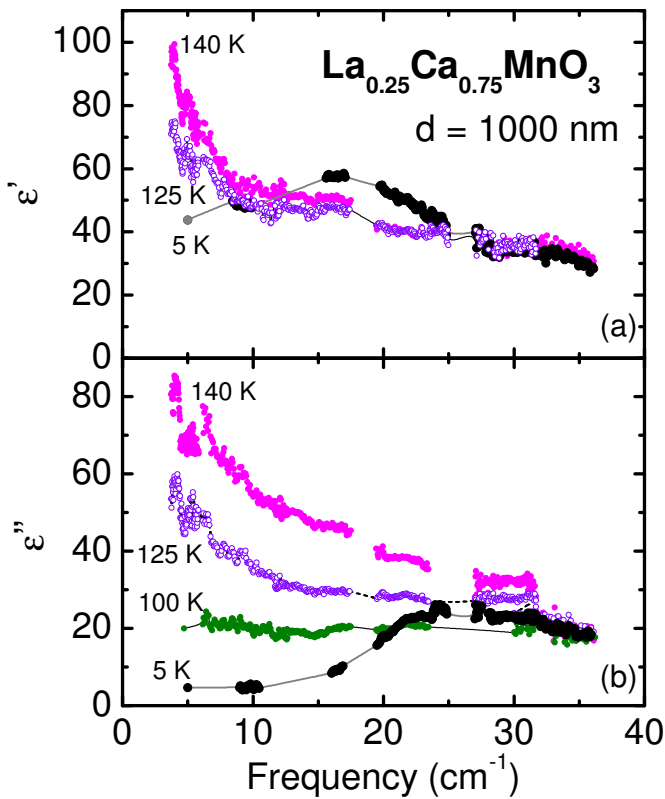}
    \caption{(Color online)  Low-frequency spectra of the (a)~real and (b)~imaginary parts of the dielectric permittivity, $\epsilon^{\prime}$ and $\epsilon^{\prime\prime}$, of  \LCMO\  directly measured in a Mach-Zehnder geometry at different temperatures as indicated. The lines are guides to the eye.
\label{fig:2}}
\end{figure}
In Fig.~\ref{fig:2} we present the pronounced relaxational behavior of the real and imaginary parts of the dielectric permittivity, $\hat{\epsilon}(\nu)$ observed at our lowest frequencies. The relaxation becomes weaker as the temperature is lowered and the dc conductivity decreases, indicating a close connection to the itinerant charge carriers. The characteristic frequency $1/(2\pi c \tau_D)$ (with $\tau_D$ the relaxation time), at which the maximum of $\epsilon^{\prime\prime}(\nu)$ is seen in a simple Debye case \cite{Jonscher83} has to be located well below our frequency range ($\nu<4$~\cm). This value is too low for the relaxation to be explained by trivial geometrical localization within the grains: from the metallic strands model the relaxation is expected at 300~\cm\ and higher \cite{Rice73}. We thus expect that the observed relaxational dispersion in \LCMO\ is intrinsic and linked to the CO process, that might be similar to the ``order-disorder'' phase transitions, for instance, in ferroeletcrics or dipole glasses (relaxors). The dynamics of such systems is commonly described by a relaxational behavior of a certain response function, with a single relaxation time or distribution of those, revealing definite temperature dependences \cite{Blink74}. Alternatively, Efremov {\it et al.} suggested that in manganites the interplay between charge and magnetism can lead to ferroelectricity with the dipole moments aligned along the diagonal between the $a$ and $b$ axes \cite{Efremov04}. The transition to such a ferroelectric phase could also be responsible for the dielectric relaxation observed in \LCMO.

\section{Conclusion}
We have investigated the conductivity and dielectric permittivity
of polycrystalline \LCMO\ in the frequency range from 4 to
700~\cm\ (quantum energies 0.4 to 90 meV) using coherent-source
THz spectroscopy, combined with FTIR measurements. In the
antiferromagnetically ordered phase at $T < T_N \approx 140$~K, a
resonance-like absorption bands appear in the spectra at $20 -
40$~\cm\ and around 100~\cm. We show that the bands are not
connected to the collective response of the charge-ordered
subsystem but are acoustic phonons which gain optical activity
when folded to the Brillouin zone center due to a structural phase
transition resulting in a fourfold superlattice along the
$a$-direction. With decreasing size of the crystallites, the
$4a$-superlattice and hence the absorption bands are gradually
suppressed while the charge and antiferromagnetic orders survive,
indicating a weak coupling of the correspondent order parameters
to the lattice. Our results evidence that all optically active
excitations of the electronically correlated phase in \LCMO\
manganite, if any, should have energies smaller than 0.4~meV.
Regardless of their microscopic origin~-- charge-modulated systems
that have correlations of the order of 0.1 to 1~eV (as
characterized by the charge gap) in general exhibit rather soft
excitations observed at microwave-, radio-, and audio-frequencies
\cite{Littlewood87,Gruner88,Vuletic06}.
At the lowest frequencies $4 - 20$~\cm\  a strong relaxation is
observed above 100~K in all samples. It may indicate an
order-disorder character of the charge-order phase transition in
\LCMO.

\section{Acknowledgements}
We acknowledge valuable discussions with  P. Arseev, D. Efremov,
N. Kovaleva, A. Mukhin, S. Tomi{\'c}, and V. Travkin. We thank G.
Untereiner for technical help. T.Z. and D.W. thank the Alexander
von Humboldt.  The work was supported by the Deutsche
Forschungsgemeinschaft (DFG) and by the RAS Program for
fundamental research ``Strongly correlated electrons in solids and
solid structures''. Further funding by the scientific programs of
the Russian Academy of Sciences and the Russian Foundation for
Basic Research (project 08-02-00757a) is acknowledged.

\end{document}